\documentclass{article}
\usepackage{cite}
\usepackage{wrapfig}
\usepackage{graphicx}
\usepackage{amssymb}
\usepackage{amsfonts}
\usepackage{amsmath}
\usepackage{longtable}
\usepackage{rotating}
\usepackage{lscape}
\usepackage{epsfig}
\usepackage{multirow}
\usepackage{placeins}
\usepackage{color}
\usepackage{float}
\usepackage{enumitem}
\usepackage{tikz}
\usetikzlibrary{arrows.meta}
\usetikzlibrary{bending}
\usepackage{amsthm}

\title{Refined $E_n$ Chern-Simons theory}

\author{M.Y. Avetisyan\footnote{E-mail: maneh.avetisyan@gmail.com},
R.L.Mkrtchyan\footnote{E-mail: mrl55@list.ru}}
\begin{document}
\maketitle

\begin{center}
  {\small {\it Yerevan Physics Institute, Yerevan, Armenia}}\\
\end{center}

\begin{abstract}
	
The partition function of refined Chern-Simons theory on 3d sphere for the exceptional $E_n$ gauge algebras is presented in terms of multiple sine functions. 
Gopakumar-Vafa (BPS) approximation is calculated and presented in the form of some refined topological string partition function. 
 
\vspace{0.2cm}

\end{abstract}
\vspace*{6pt}

\noindent
PACS: 11.15.Yc; 11.25.-w

\label{sec:intro}

\section{Introduction}

Refined Chern-Simons (CS) theory was suggested in \cite{AS11,AS12a} on the basis of Macdonald's deformation of different objects in the theory of simple Lie algebras, particularly of modular matrices $S$ and $T$. It was studied for A and D type gauge algebras and the duality with refined topological string theory \cite{N02,HIV,IKV07} was shown in the simplest case of the theory on $S^3$ manifold. The closed expression for partition function of refined CS theory on $S^3$ for an arbitrary gauge algebra was suggested in our work \cite {AM21}, based on the refined Kac-Peterson identity \cite{KP} for the determinant of symmetrized Cartan matrix. According to the general approach developed in \cite{M13} this universal-type expression allows one to transform the partition function into a Gopakumar-Vafa (BPS) type \cite{GV1} expression for the partition function of a topological string, establishing the duality between the two theories. The duality 
was extended to B and C type gauge algebras in \cite{AM22}. In the present paper we consider the case of the exceptional E type algebras, and present partition functions of corresponding refined Chern-Simons theories in the form of partition functions of some unidentified refined topological string theory.

\section{ Refinement of a E type CS theory}

Let's recall the universal expression for the free energy of the CS theory in terms of Vogel's universal parameters (see \cite{MV,M13} for the formula below and the definition of Vogel's universal parameters $\alpha, \beta, \gamma$):
\begin{equation}\label{u1}
	\mathcal{F}=\frac{1}{4} \int_{R_{+}} \frac{dx}{x} \frac{sh(x(t-\delta))}{sh(x t) sh(x \delta)} f(2x),
\end{equation}
where
\begin{equation}
	f(x)=\frac{sh(x \frac{\alpha-2t}{4}) sh(x \frac{\beta-2t}{4}) sh(x \frac{\gamma-2t}{4})}{sh(x\frac{\alpha}{4}) sh(x\frac{\beta}{4}) sh(x\frac{\gamma}{4})}
\end{equation}
is the universal quantum dimension of the simple Lie algebras,
$ t=\alpha+\beta+\gamma,$
and $\delta=t+\kappa$, $\kappa$ is the coupling constant of the CS theory. Different sets of Vogel's parameters $\alpha, \beta, \gamma$ correspond to different (all) simple Lie algebras according 
to Vogel's table \cite{MV}. 

Now consider (\ref{u1}) as a function of the quintuple of independent parameters:  $(\alpha, \beta,\gamma,t,\kappa)$. Following \cite{KS} let us introduce the followin deformation of the parameters:  $(\alpha, \beta,\gamma,t,\kappa)\rightarrow (\alpha, y\beta, y\gamma, y t, \kappa)$ where the refinement parameter $y$ was introduced. In \cite{KS} it is stated thet then we get the 
partition function of the refined CS with refinement parameter $y$. In \cite{AM21}, this statement was extended on the E type algebras.

Next, we take a look at the E type algebras. Recall the equation of the exceptional line: $\gamma=2(\alpha+\beta)$. 
We parametrize it in the following way: $\alpha \rightarrow z, \beta \rightarrow 1-z, \gamma \rightarrow 2$, so that $t=\alpha+\beta+\gamma=3$. 
The $E_n, n=6,7,8$ algebras correspond to $z= -\frac{1}{2}, -\frac{1}{3}, -\frac{1}{5}$, respectively. 

Our first goal is to express the free energy in terms of multiple sine functions. The second one consists of the derivation of the "perturbative" (from the point of view of dual string) part of t
his partition function. 

Let us start with the integrand of (\ref{u1}) for the $(\alpha, y\beta, y\gamma, y t, \kappa)$ set of the parameters:
\begin{equation}
	\frac{1}{2} \frac{ch(x y) sh(x (3 y-\delta)) sh(\frac{x}{2} y (z+5) sh(\frac{x}{2}(z-6 y))}{sh(x \delta) sh(3 x y) sh(\frac{x}{2}z) sh(\frac{x}{2} y (1-z))}
\end{equation}

which can be rewritten in the following form:
\begin{multline}\label{4}
	-\frac{8 e^{x \delta} e^{3 x y} e^{\frac{1}{2} (x z)} e^{\frac{1}{2} x y (1-z)} ch(x y) sh (x (3 y-\delta)) sh(\frac{1}{2} x y (z+5)) sh(\frac{1}{2} x (z-6 y))}
	{(e^{2 x \delta}-1) (e^{6 x y}-1) (e^{-x z}-1) ( e^{x y (1-z)}-1)} 
\end{multline}

We then expand the numerator of (\ref{4}) into exponents obtaining:
\begin{multline}\label{5}
	-\frac{1}{2} \frac{e^{x (2 \delta+5 y-z)}+ e^{x (2 \delta+7 y-z)}+ e^{x (2 \delta-y z-6 y)}+ e^{x (2 \delta-y z-4 y)}- e^{x (2 \delta-y z-z)} }{(e^{2 x \delta}-1) (e^{6 x y}-1) (e^{-x z}-1) ( e^{x y (1-z)}-1)}- \\ 
	- \frac{e^{2 \delta x-x y z+2 x y-x z}
		- e^{x (2 \delta-y)}-
		e^{x (2 \delta+y)}- e^{x (11 y-z)}- e^{x (13 y-z)}- e^{-x y (z-2)}- 
		e^{-x y z}}{(e^{2 x \delta}-1) (e^{6 x y}-1) (e^{-x z}-1) ( e^{x y (1-z)}-1)}+ \\
	+\frac{ e^{-x (y z-8 y+z)}+ e^{-x (y z-6 y+z)}+ e^{5 x y} + e^{7 x y}}{(e^{2 x \delta}-1) (e^{6 x y}-1) (e^{-x z}-1) ( e^{x y (1-z)}-1)}
\end{multline}

Integral of (\ref{5}) can be presented in terms of the multiple sine functions $S_r(z|\underline{\omega})$:

\begin{equation}\label{Sint}
		\ln S_r(z|\underline{\omega})\sim 
		(-1)^r \int_{R_+}\frac{dx}{x}	\frac{e^{zx}}{\prod_{k=1}^r(e^{\omega_i x}-1)}\,,
	\end{equation}
(see definitions and the set of identities in \cite{AM22}), so that the partition function rewrites as a
product of quadruple sine functions. 

Specifically, the corresponding partition function will be the square root of the following expression of quadruple sines:
\begin{multline} \label{6}
	\frac{S_4(2\delta-z(1+y) \vert \omega)}{S_4( 7y \vert \omega)} \cdot \frac{S_4(2\delta+2y-z(1+y) \vert \omega)}{S_4( 5y \vert \omega)} \times \\
	\frac{S_4(-z y \vert \omega)}{S_4(2\delta+7y-z \vert \omega)} \cdot \frac{S_4(2y-z y \vert \omega)}{S_4(2\delta+5y-z \vert \omega)}
	\times \\
	\frac{S_4(13 y-z\vert \omega)}{S_4(2\delta-6y-z y \vert \omega)} \cdot \frac{S_4(11 y-z \vert \omega)}{S_4(2\delta-4 y-z y \vert \omega)} \times \\
	\frac{S_4(2\delta-y \vert \omega)}{S_4(8 y-z(1+y) \vert \omega)} \cdot \frac{S_4(2\delta+y \vert \omega)}{S_4(6y-z(1+y) \vert \omega)}
\end{multline}
where $\omega$ stands for the set of $2\delta, 6 y, y(1-z), -z$ parameters.
Notice, that for each of the $S_4(\mathcal{X} \vert \omega)$ function there is one with $S_4(\underline{\omega}-\mathcal{X} \vert \omega)$ ($\underline{\omega}$ is the sum of all parameters $\omega$). Employing the identities 
for the multiple sine functions, we simplify (\ref{6}) into
\begin{multline} \label{7}
	\frac{S_4(-y z\vert \omega)\cdot S_4(2 y-y z\vert \omega)\cdot S_4(13y - z\vert \omega)\cdot S_4(11 y - z\vert \omega)}{S_4(7y \vert \omega)\cdot S_4(5y \vert \omega)\cdot 
		S_4(8 y-z(1+y) \vert \omega)\cdot S_4(6y-z(1+y) \vert \omega)}
\end{multline}

With the same set of the identities employed (\ref{7}) reduces into the following expression of triple sine functions:
\begin{multline}\label{8}
	\frac{S_3(7y-z\vert 6y, y(1-z), -z)}{S_3(7y-z\vert 2\delta, y(1-z), -z)} \cdot \frac{S_3(5y-z\vert 6y, y(1-z), -z)}{S_3(5y-z\vert 2 \delta, y(1-z), -z)} \times \\
	\frac{S_3(y\vert 2\delta, y(1-z), -z)}{S_3(y\vert 6y, y(1-z), -z)} \cdot \frac{S_3(2y-z(1+y)\vert 2\delta, y(1-z), -z)}{S_3(5y\vert 6y, y(1-z), -z)}
\end{multline}
Now we transform this expression to get the multiple sines with the following set of parameters: $(2\delta, y(1-z),-z)$, and finally we have for partition function
\begin{multline} \label{9}
	Z=
	S_2(-yz\vert y(1-z),-z)\cdot\frac{S_1(y\vert 2\delta)}{S_1(5y\vert 6y)}\times \\
	\frac{S_3(2y-y z\vert 2\delta,y(1-z),-z)\cdot S_3(y-z\vert 2\delta,y(1-z),-z)}{S_3(7y- z\vert 2\delta,y(1-z),-z)\cdot S_3(5y-z\vert 2\delta,y(1-z),-z)}
\end{multline}
At $y=1$ this expression coincides with the one derived in \cite{M20}.

This is an exact expression for the partition function similar to those previously studied - pure  A, B, C, D, E, F, G or refined A, B, C, D CS theories. In these cases the corresponding expressions
allow to establish duality with the (refined) topological strings, to calculate non-perturbative corrections to string partition functions, etc. In the next section we calculate the  Gopakumar-Vafa  (BPS) approximation of this refined E type partition function, which appears to be similar to the refined topological string theories, although the exact string theory remains unidentified. 

\section{Perturbative poles approximation of the partition function}

Closing the integration contour in the integral representation (\ref{Sint}) of multiple sines in the upper semiplane (or, the same, making use of the (2.14) formula in \cite{KM}) we can rewrite $ln(Z)$ as a sum of the poles of the integrands of (\ref{Sint}) in the upper semiplane. 
We are interested only in the perturbative part, so the sum over the perturbative poles \cite{M13,KM} (i.e. those, corresponding to denominator $(exp(2\delta)-1)$) will be:
\begin{quotation} \tiny
\begin{multline}\label{10}
	ln(Z)\cong \sum_{n=1}^{\infty} \frac{
		e^{\frac{\pi i n (5 y-z)}{\delta}}+
		e^{\frac{\pi i n (7 y-z)}{\delta}}-e^{\frac{\pi i n (2 y-z y)}{\delta}}-e^{\frac{\pi i n (y-z) }{\delta}}}
	{n \left(e^{-\frac{\pi i n z}{\delta}}-1\right) \left(e^{\frac {\pi i n y (1-z)}{\delta}}-1\right)} =\\
	\sum_{n=1}^{\infty} \frac{-e^{\frac{2y \pi i n}{2\delta}} 
		\left(e^{-\frac{(-y-z+y z) \pi i n}{2 \delta}}+e^{\frac{(-y-z+y z) \pi i n}{2 \delta}} \right)+
		e^{\frac{(11y-z+y z) \pi i n}{2 \delta}} 
		\left(e^{-\frac{-2y \pi i n}{2 \delta}} + e^{-\frac{2y \pi i n}{2 \delta}} \right)}
	{n \left(-e^{-\frac{y(1-z) \pi i n}{2 \delta}}+e^{\frac{y(1-z) \pi i n}{2 \delta}}\right) \left(e^{-\frac{z \pi i n}{2 \delta}}-e^{\frac{z \pi i n}{2 \delta}}\right)}
\end{multline}
\end{quotation}

One can expect to observe the correspondence of (\ref{10}) with the refined topological string partition function: 

\begin{quotation} \tiny
	\begin{eqnarray} \label{ztop}
		&\ln Z_{top} \cong \\ \nonumber
		&-\sum_{C\in H_2(X,\mathbb{Z})} \sum_{n=1}^{\infty}\sum_{j_L,j_R} \frac{(-1)^{2j_L+2j_R}N^C_{j_L,j_R}((qt)^{-nj_L}+...+(qt)^{nj_L})((\frac{q}{t})^{-nj_R}+...+(\frac{q}{t})^{nj_R})}{n(q^{\frac{n}{2}}-q^{-\frac{n}{2}})(t^{\frac{n}{2}}-t^{-\frac{n}{2}})} e^{-nT_C}
	\end{eqnarray}
\end{quotation}

as it is done in \cite{M20} at $y=1$, but it appears that this expression is not general enough. 
Indeed, the choice of $q$ and $t$ is restricted to:
$$q=e^{-\frac{\pi i z}{\delta}}, t=e^{\frac{\pi i y (1-z)}{\delta}} $$
or vice versa.
With this correspondence, the $q t$ and $\frac{q}{t}$ are $e^{-\frac{\pi i (y (z-1)+z)}{\delta}}$ and $e^{\frac{\pi i (y (z-1)-z)}{\delta}}$, respectively, which 
obviously do not allow one to interpret (\ref{10}) as (\ref{ztop}). The possible explanation is that (\ref{ztop})
is the most general form of orientable refined topological string, only, not including non-orientable surfaces. 
 The non-orientable (i.e. B, C, D types) refined topological strings partition functions appeared in \cite{AM22}. One has to construct the general form of them and compare with (\ref{10}), to finally obtain an interpretation of different contributions in (\ref{10}) in terms of surfaces of different types.

\section*{Acknowledgments} 
	
We are indebted to A.Mironov for discussion of refined topological string theories, and to the organizers of SQS'22 conference for invitation. 
The work of MA was fulfilled within the Regional Doctoral Program on Theoretical and Experimental Particle Physics  sponsored by VolkswagenStiftung. MA and RM are partially supported by the Science Committee of the Ministry of Science and Education of the Republic of Armenia under contract 21AG-1C060. The work of MA is partially supported by ANSEF grant PS-mathph-2697.

\end{document}